\documentclass[aps,prl,twocolumn,floatfix, preprintnumbers]{revtex4}
\usepackage[pdftex]{graphicx}
\usepackage[mathscr]{eucal}
\usepackage[pdftex]{hyperref}
\usepackage{amsmath,amssymb,bm}

\newif\ifarxiv
\arxivfalse

\begin		{document}

\def\Nfour	{\mathcal N\,{=}\,4}

\def\Nc		{N_{\rm c}}

\def\gauge	{\xi}
\def\half	{{\textstyle \frac 12}}

\def\section	#1{\quad\textit{#1}.---}

\preprint{MIT-CTP 4192}

\title
    {
    Holography and colliding gravitational shock waves in asymptotically AdS$_5$ spacetime
    }

\author{Paul~M.~Chesler}
\affiliation
    {Department of Physics, MIT, Cambridge, MA 02139, USA}
\email{pchesler@mit.edu}

\author{Laurence~G.~Yaffe}
\affiliation
    {Department of Physics, University of Washington, Seattle, WA 98195, USA}
\email{lgy@uw.edu}

\date{\today}

\begin{abstract}
Using holography, we study the collision of planar shock waves
in strongly coupled $\Nfour$ supersymmetric Yang-Mills theory.
This requires the numerical solution of a dual gravitational
initial value problem in asymptotically anti-de Sitter spacetime.
\end{abstract}



\maketitle
\parskip	2pt plus 1pt minus 1pt

\noindent \section{{Introduction}}
The recognition that the quark-gluon plasma (QGP) produced in relativistic
heavy ion collisions is strongly coupled \cite{Shuryak:2004cy},
combined with the advent of gauge/gravity duality (or ``holography'')
\cite{Maldacena:1997re,Witten:1998qj},
has prompted much work exploring both equilibrium and non-equilibrium
properties of strongly coupled $\Nfour$ supersymmetric Yang-Mills
theory (SYM), which may be viewed as a theoretically tractable
toy model for real QGP.
Multiple authors have discussed collisions of
infinitely extended planar shock waves in SYM, which
may be viewed as instructive caricatures of collisions of
large, highly Lorentz-contracted nuclei.
In the dual description of strongly coupled (and large $\Nc$) SYM,
this becomes a problem of colliding gravitational shock waves
in asymptotically anti-de Sitter (AdS$_5$) spacetime.
Previous work has examined qualitative properties and trapped surfaces
\cite{%
Albacete:2009ji,
Kovchegov:2009du,
Gubser:2009sx,
Lin:2009pn},
possible early time behavior
\cite{Kovchegov:2007pq,
Grumiller:2008va,
Beuf:2009cx},
and expected late time asymptotics
\cite{%
Janik:2005zt,
Janik:2006gp}.
As no analytic solution is known for this gravitational problem,
solving the gravitational initial value problem numerically is
the only way to obtain quantitative results which properly
connect early and late time behavior.
In this letter, we report the results of such a calculation, and
examine the evolution of the post-collision stress-energy tensor.

Unlike previous work considering
singular shocks with vanishing thickness 
\cite{%
Grumiller:2008va,
Kovchegov:2009du},
or shocks driven by non-vanishing sources
in the bulk 
\cite{%
Kovchegov:2009du,
Gubser:2009sx},
we choose to study planar gravitational ``shocks'' which are regular,
non-singular, source-less solutions to Einstein's equations.
Equivalently, we study the evolution
of initial states in SYM with finite energy density concentrated on
two planar sheets of finite thickness (and Gaussian profile),
propagating toward each other at the speed of light.
The dual description only involves gravity in
asymptotically AdS$_5$ spacetime;
the complementary 5D internal manifold plays no role and may be ignored.
Consequently, our results apply to all strongly coupled 4D
conformal gauge theories with a pure gravitational dual description.

\noindent \section{{Gravitational description}}
Diffeomorphism invariance plus translation invariance in two spatial
directions allows one to write the 5D spacetime metric in the form
\begin{equation}
    ds^2 = -A \, dv^2
    + \Sigma^2 \left[ e^B d{\bm x}_\perp^2 + e^{-2B} dz^2 \right]
    + 2 dv \, (dr + F dz) \,,
\label{eq:ansatz}
\end{equation}
where $A$, $B$, $\Sigma$, and $F$ are all functions of the bulk radial
coordinate $r$, time $v$, and longitudinal coordinate $z$.
We employ generalized infalling Eddington-Finkelstein coordinates.
Lines along which all coordinates except $r$ are constant are
infalling radial null geodesics;
the radial coordinate $r$ is an affine parameter
along these geodesics.
At the boundary, located at $r = \infty$, $v$ coincides with time in the dual 
quantum field theory.
The geometry in the bulk at $v \ge 0$ is the causal future of
$v = 0$ on the boundary.  
The ansatz (\ref{eq:ansatz}) is invariant under the residual
diffeomorphism $r \to r + \gauge$, with $\gauge$
an arbitrary function of $v$ and $z$.

For a metric of the form (\ref{eq:ansatz}),
Einstein's equations
(with cosmological constant $\Lambda \equiv -6$) imply
\begin{subequations}
\begin{align}
    0 &= \Sigma'' + \half (B')^2 \, \Sigma \,,
\label{eq:Sigmaeqn}
\\[4pt]
    0 &= \Sigma^2 \left[ F'' -2 (d_3 B)' -3 B' d_3 B \right]
	+4 \Sigma' d_3 \Sigma \,,
\nonumber\\ &\qquad{}
	-\Sigma \left[ 3\Sigma' F' + 4 (d_3 \Sigma)' + 6 B' d_3 \Sigma \right],
\label{eq:Feqn}
\\[4pt]
    0 &= \Sigma^4 \left[ A'' + 3 B' d_+ B + 4 \right]
	-12 \Sigma^2 \Sigma' d_+ \Sigma
\nonumber\\ &\quad{}
    + e^{2B} \bigl\{
		\Sigma^2 \left[ \half (F')^2 {-} \tfrac 72 (d_3 B)^2 {-}  2 d_3^2 B \right]
\nonumber\\ &\qquad{}
		+ 2 (d_3 \Sigma)^2
		- 4 \Sigma \left[ 2 (d_3 B)d_3 \Sigma + d_3^2 \Sigma \right]
	    \bigr\} \,,
\label{eq:Aeqn}
\\[4pt]
    0 &= 6 \Sigma^3 (d_+ \Sigma)'
	+ 12 \Sigma^2 (\Sigma' d_+ \Sigma - \Sigma^2)
	- e^{2B} \left\{ \vphantom{\big|} 2(d_3\Sigma)^2 \right.
\nonumber\\ &\quad{}
	+ \Sigma^2 \!\left[
		\tfrac 12 (F')^2
		{+} (d_3 F)'
		{+} 2F' d_3 B
		{-} \tfrac 72 (d_3 B)^2
		{-} 2d_3^2 B
	    \right]
\nonumber\\ &\qquad\left.{}
    + \Sigma \left[ (F' {-} 8 d_3 B)\, d_3\Sigma - 4 d_3^2 \Sigma \right]
	    \right\}.
\label{eq:Sigmadot}
\\[4pt]
    0 &= 6 \Sigma^4 (d_+ B)' + 9 \Sigma^3 (\Sigma' d_+ B + B' d_+ \Sigma)
\nonumber\\ &\quad{}
    + e^{2B} \bigl\{
	    \Sigma^2 [ (F')^2 {+} 2 (d_3 F)' {+} F' d_3 B {-} (d_3 B)^2 {-} d_3^2 B ]
\nonumber\\ &\qquad{}
	    + 4 (d_3 \Sigma)^2
	    - \Sigma \left[
		    (4 F' {+} d_3 B)\, d_3 \Sigma + 2 d_3^2 \Sigma
		\right]
	    \bigr\} \,,
\label{eq:Beqn}
\\[4pt]
    0 &= 6 \Sigma^2 d_+^2 \Sigma - 3 \Sigma^2 A' d_+\Sigma 
	+ 3 \Sigma^3 (d_+B)^2
\nonumber\\ &\quad{}
    - e^{2B} \left\{\vphantom{\big|}
     (d_3\Sigma + 2 \Sigma d_3 B)(2 d_+ F + d_3 A) 
    \right.
\nonumber\\ &\qquad\left.\vphantom{\big|}{}
    + \Sigma \left[ 2d_3 (d_+ F) + d_3^2 A \right]
    \right\},
\label{eq:xtra2}
\\[4pt]
    0 &= \Sigma \left [
      2 d_+ (d_3 \Sigma)
    + 2 d_3 (d_+ \Sigma)
    + 3 F' d_+\Sigma \right]
\nonumber\\ &\quad{}
    + \Sigma^2 \left[
	  d_+ (F') 
	+ d_3 (A')
	+ 4d_3(d_+ B)
	- 2d_+(d_3 B)
	\right]
\nonumber\\ &\quad{}
	+ 3 \Sigma \left( \Sigma d_3 B + 2 d_3\Sigma \right) d_+ B
	- 4 (d_3 \Sigma) d_+\Sigma \,,
\label{eq:xtra1}
\end{align}\label{eq:einseqns}
\end{subequations}
where, for any function $h(v,z,r)$, $h' \equiv \partial_r h$ and
\begin{equation}
    d_+ h \equiv \partial_v h + \half A\, \partial_r h \,,\quad
    d_3 h \equiv \partial_z h - F\, \partial_r h \,.
\end{equation}
Note that $h'$ is a directional derivative along infalling radial null
geodesics, $d_+ h$ is a derivative along outgoing radial null geodesics,
and $d_3 h$ is a derivative in the longitudinal direction orthogonal to
both radial geodesics.

Near the boundary, Einstein's equations may be solved with a power series
in $r$.
Solutions with flat Minkowski boundary geometry have the form
\begin{subequations}
\begin{align}
    A &= r^2 \Big[
		    1 + \frac {2 \gauge} r
		    + \frac {\gauge^2 {-} 2 \partial_v \gauge}{r^2}
		    + \frac {a_4}{r^{4}} + O(r^{-5})
		\Big] \,,
\label{eq:Aexp}
\\
    F &= \partial_z \gauge + \frac {f_2}{r^2} + O(r^{-3})\,.
\label{eq:Fexp}
\\
    B &= \frac{b_4}{r^{4}} + O(r^{-5})\,,
\label{eq:Bexp}
\\
    \Sigma &= r + \gauge + O(r^{-7})\,,
\label{eq:Sigmaexp}
\end{align}\label{eq:asymptotics}%
\end{subequations}
The coefficient $\gauge$ is a gauge dependent parameter which 
encodes the residual diffeomorphism invariance of the metric.  
The coefficients $a_4$, $b_4$ and $f_2$
are sensitive to the entire bulk geometry,
but must satisfy
\begin{equation}
\label{eq:boundaryevolution}
\partial_v a_{4} = -{\tfrac{4}{3}}\, \partial_z f_{2} \,, \quad
\partial_v f_{2} = -\partial_z ({\tfrac{1}{4}}a_{4} +2 b_{4})\,.
\end{equation}
These coefficients
contain the information which,
under the holographic mapping of gauge/gravity duality,
determines the field theory stress-energy tensor $T^{\mu \nu}$ \cite{deHaro:2000xn}.
Defining
$\mathcal {E}    \equiv  \tfrac {2\pi^2}{\Nc^{2\vphantom{)}}} \, T^{00}$, 
$\mathcal {P}_\perp \equiv \tfrac {2\pi^2}{\Nc^{2\vphantom{)}}} \, T^{\perp \perp}$, 
$\mathcal {S}    \equiv \tfrac {2\pi^2}{\Nc^{2\vphantom{)}}} \, T^{0z}$,
and $\mathcal {P}_\| \equiv \tfrac {2\pi^2}{\Nc^{2\vphantom{)}}} \, T^{zz}$,
one finds
\begin{subequations}%
\label{eq:holorenorm}
\begin{align}
  \mathcal {E} &= - \tfrac 34 a_4 \,,
& \mathcal {P}_\perp &= - \tfrac 14 a_4 + b_4 \,,
\\
  \mathcal {S} &= -f_2 \,,
& \mathcal {P}_\| &= -\tfrac 14 a_4 - 2 b_4 \,.
\end{align}%
\end{subequations}
Eqs.~(\ref{eq:boundaryevolution}) and (\ref{eq:holorenorm})
imply $\partial_\mu T^{\mu\nu} = 0$ and $T^{\mu}_{\ \mu} = 0$.

\noindent \section{{Numerics overview}}
Our equations (\ref{eq:einseqns}) have a natural nested linear structure
which is extremely helpful in solving for the fields and their 
time derivatives on each $v = {\rm const.}$ null slice.  
Given $B$,
Eq.~(\ref{eq:Sigmaeqn}) may be integrated in $r$ to find $\Sigma$.
With $B$ and $\Sigma$ known, Eq.~(\ref{eq:Feqn}) may be integrated to find $F$.
With 
$B$, $\Sigma$ and $F$ known, Eq.~(\ref{eq:Sigmadot}) may be integrated to find
$d_+ \Sigma$.  With $B$, $\Sigma$, $F$ and $d_+ \Sigma$ known,  Eq.~(\ref{eq:Beqn})
may be integrated to find $d_+B$.  Last, with $B$, $\Sigma$, $F$, $d_+ \Sigma$ and $d_+ B$
known, Eq.~(\ref{eq:Aeqn}) may be integrated to find $A$.
At this point,
one can compute the field velocity 
$\partial_v B = d_+B - \frac{1}{2} A B'$, evolve $B$
forward in time to the next time step,
and repeat the process.

In this scheme, each nested equation
is a linear ODE for the field being determined,
and may be integrated in $r$ at fixed $v$ and $z$.
The requisite radial boundary 
conditions follow from the asymptotic expansions (\ref{eq:asymptotics}).
Consequently, the initial data required to solve Einstein's equations 
consist of the function $B$ plus the expansion coefficients $a_4$ and $f_2$
--- all specified at some constant $v$ ---
and the gauge parameter $\gauge$ specified at all times.  
Values of $a_4$ and $f_2$ on future time slices, needed as boundary
conditions for the radial equations, are determined by integrating
the continuity relations (\ref{eq:boundaryevolution}) forward in time.

Eqs.~(\ref{eq:xtra2}) and (\ref{eq:xtra1}) are only needed
when deriving the series expansions (\ref{eq:asymptotics}) and
the continuity conditions (\ref{eq:boundaryevolution}).
In this scheme, they are
effectively implemented as boundary conditions.
Indeed, the Bianchi identities imply that Eqs.~(\ref{eq:xtra2}) and (\ref{eq:xtra1})
are boundary constraints;
if they hold at one value of $r$ then the other
Einstein equations guarantee that they hold at all values of $r$.

An important practical matter 
is fixing the computational domain in $r$.  If an event horizon exists, then
one may excise the geometry inside the horizon, as this region is causally
disconnected from the outside geometry.  Moreover, one 
\emph{must} excise the geometry to avoid singularities behind the horizon
\cite{Anninos:1994dj}.
To perform the excision,
we identify the location of an apparent horizon (an outermost
marginally trapped surface) which, if it exists, must lie inside an
event horizon \cite{Wald:1984rg}.  
For the initial conditions
discussed in the next section, the apparent horizon 
always exists --- even before the collision --- and has the topology of a plane.
Hence,
one may fix the residual diffeomorphism invariance 
by requiring the apparent horizon position to lie at a fixed radial
position, $r = 1$.  
The defining conditions for the apparent horizon then imply that
fields at $r = 1$ must satisfy 
\begin{equation}
\label{eq:horizon}
0 = 3 \Sigma^2 \, d_+ \Sigma -
 \partial_z (F \, \Sigma \, e^{2B})
  + {\textstyle \frac{3}{2}} F^2 \, \Sigma' e^{2B},
\end{equation}
which is implemented as a boundary condition to determine $\gauge$
and its evolution.
Horizon excision is performed by restricting the 
computational domain to $r \in [1,\infty]$.

Another issue is the presence of a singular point at
$r = \infty$ in the equations (\ref{eq:einseqns}).
To handle this, we
discretize Einstein's equations
using pseudospectral methods \cite{Boyd:2001}.
We represent the
radial dependence of all functions as a series in Chebyshev polynomials 
and
the $z$-dependence as a Fourier series, so the $z$-direction
is periodically compactified.  With these basis functions,
the computational domain may extend all the way to $r = \infty$,
where boundary conditions can be directly imposed.

\noindent \section{{Initial data}}
We want our initial data to describe two well-separated planar shocks,
with finite thickness and energy density, 
moving toward each other.
In Fefferman-Graham coordinates,
an analytic solution describing a single planar shock moving in the
$\mp z$ direction may be easily found
and reads \cite{Janik:2005zt}, 
\begin{equation}
    ds^2 =
    r^2 [ -d  x_{+} d x_{-} + d \bm x^2_{\perp} ]
    + \frac 1{r^{2}} \, [ {d r^2} + h(x_{\pm})\, dx_{\pm}^2 ]\,,
\label{eq:FGshock}
\end{equation}
with $x_\pm \equiv t \pm z$, and
$h$ an arbitrary function which we choose
to be a Gaussian with width $w$ and amplitude $\mu^3$,%
\begin{equation}
    h(x_\pm) \equiv \mu^3\, (2\pi w^2)^{-1/2} \, e^{-\frac 12 x_\pm^2 / w^2} 
    \,.
\end{equation}
The energy per unit area of the shock is
$\mu^3 (\Nc^2/2\pi^2)$.
If the shock profile $h$ has compact support,
then a superposition of right and left moving shocks
solves Einstein's equations at early times when the incoming shocks have
disjoint support.
Although this is not exactly true for our Gaussian profiles,
the residual error in Einstein's equations is negligible
when the separation of the incoming shocks is more than a few times the
shock width.

To find the initial data relevant for our metric ansatz (\ref{eq:ansatz}),
we solve (numerically) for the diffeomorphism
transforming the single shock metric (\ref{eq:FGshock}) from Feffer\-man-Graham 
to Eddington-Finkelstein coordinates.
In particular, we compute the anisotropy
function $B_{\pm}$ for each shock and
sum the result,
$B = B_+ + B_-$.
We choose the initial time $v_0$ so the incoming shocks are well separated 
and the $B_{\pm}$ negligibly overlap above the apparent horizon.  The functions
$a_4$ and $f_2$ may be found analytically,%
\begin{equation}
a_4 = -{\textstyle \frac{4}{3}} \left [h(v_0 {+} z) {+} h(v_0 {-} z) \right], \ 
f_2 = h(v_0 {+} z) {-} h(v_0 {-} z).
\end{equation}

A complication with this initial data is that
the metric functions $A$ and $F$ become very large deep in the bulk,
degrading convergence of their spectral representations.
To ameliorate the problem, we slightly modify the initial data,
subtracting from $a_4$ a small positive constant $\delta$.
This introduces a small background energy density in the dual quantum theory.
Increasing $\delta$ causes the regions with
rapid variations in the metric to be pushed inside the apparent horizon,
out of the computational domain.

We chose a width $w = 0.75/\mu$ for our shocks.
The initial separation of the shocks
is $\Delta z = 6.2/\mu$.  We chose $\delta = 0.014\, \mu^4$,
corresponding to a background energy density 50 times smaller than the peak energy density of the shocks.
We evolve the system for a total time
equal to the inverse of the temperature 
associated with the background energy density,
$T_{\rm bkgd} = 0.11\, \mu$.

\begin{figure}
\includegraphics[scale=0.35]{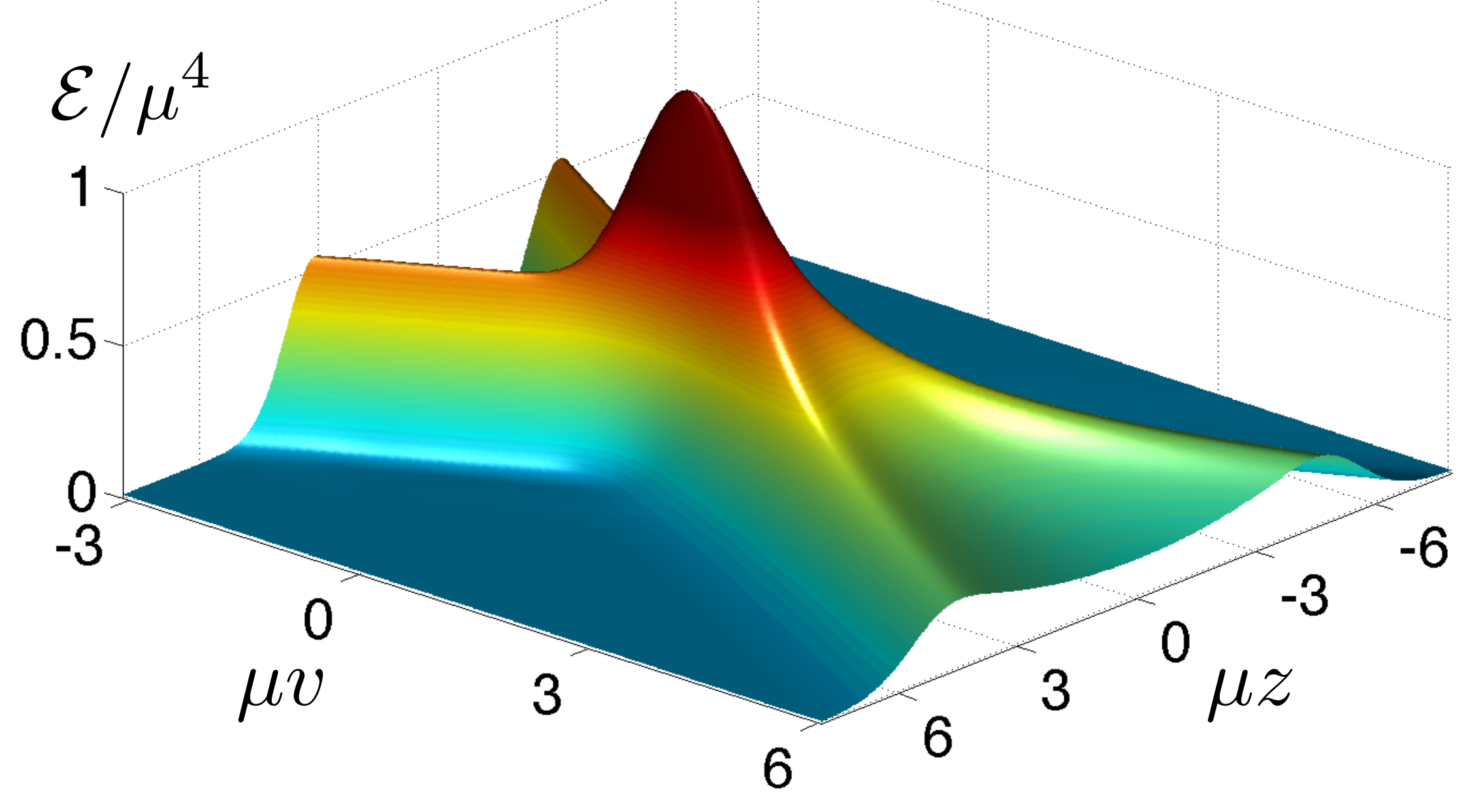}%
\caption{Energy density $\mathcal E/ \mu^4$ 
    as a function of time $v$ and longitudinal coordinate $z$.
    \label{fig:eden}}
\end{figure}

\begin{figure}
\includegraphics[scale=0.35]{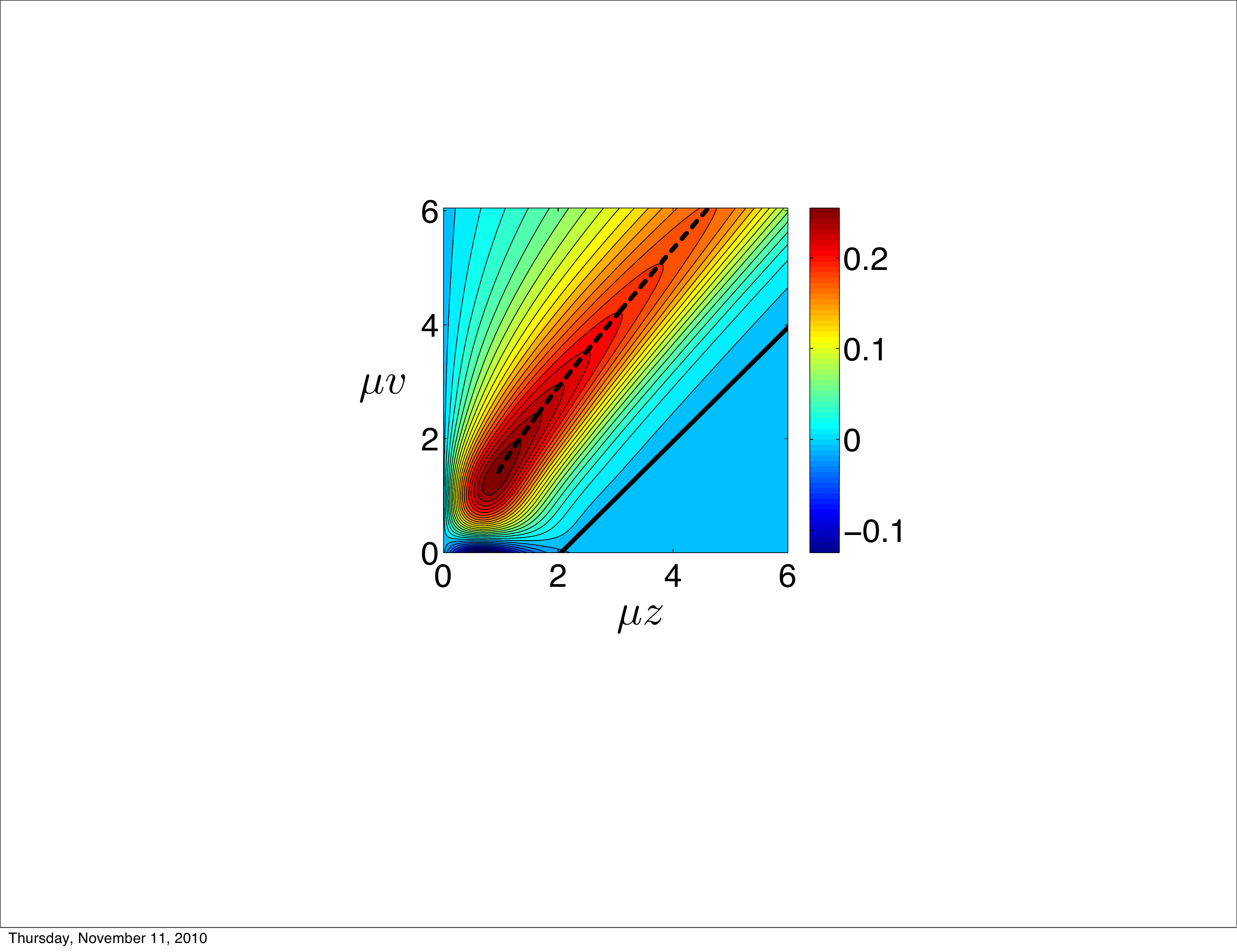}%
\vspace*{-5pt}
\caption{Energy flux $\mathcal  S/ \mu^4$ 
    as a function of time $v$ and longitudinal coordinate $z$.
    \label{fig:flux}}
\end{figure}

\begin{figure}
\includegraphics[scale=0.34]{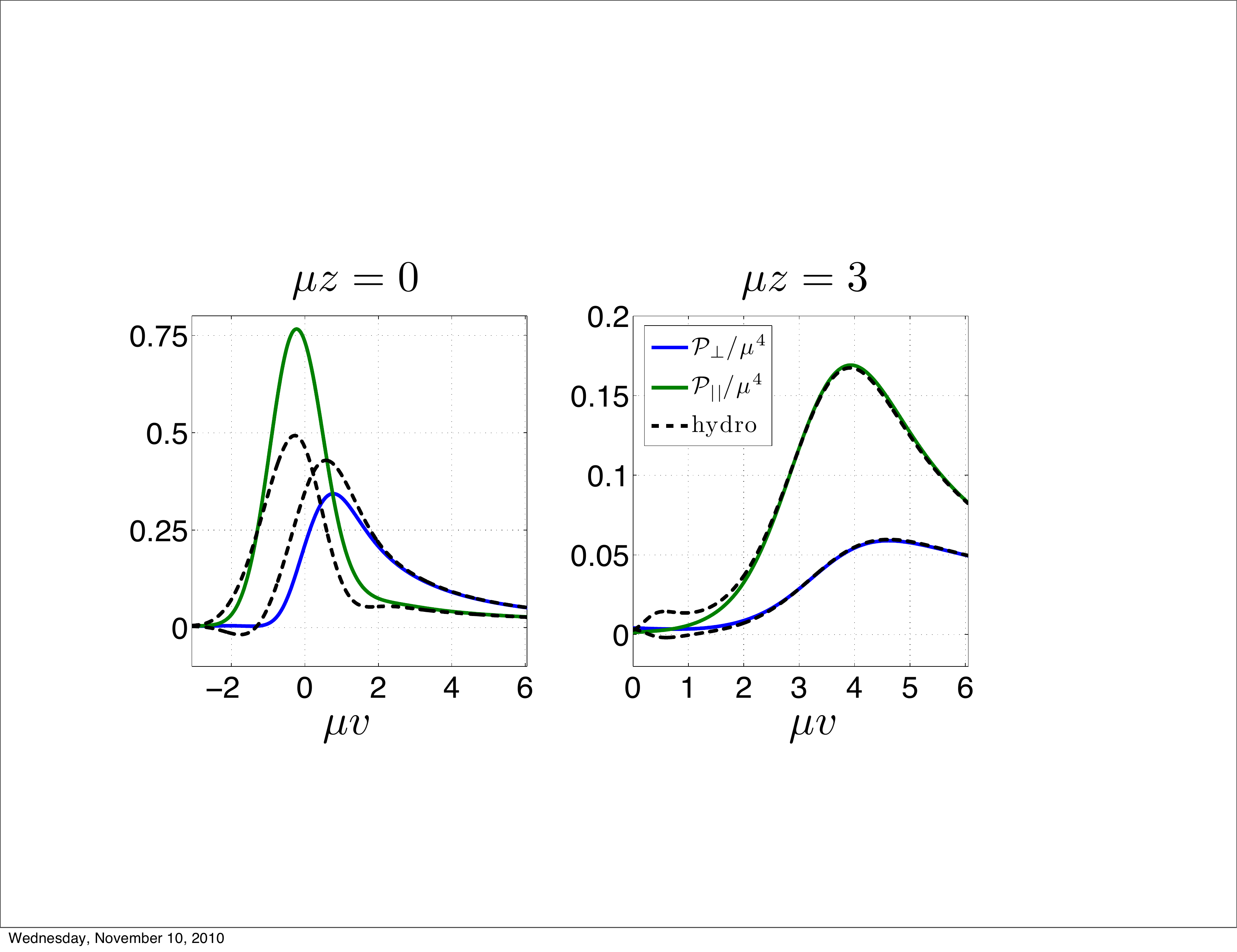}%
\caption{Longitudinal and transverse pressure as a function of time $v$,
    at $z=0$ and $z = 3/\mu$.
    Also shown for comparison are the pressures predicted by
    the viscous hydrodynamic constitutive relations.
    \label{fig:pressures}}
    \vspace*{-10pt}
\end{figure}

\noindent \section{{Results and discussion}}
Figure~\ref{fig:eden} shows the energy density $\mathcal E$ as a function of
time $v$ and longitudinal position $z$.
On the left, one sees two incoming shocks propagating toward
each other at the speed of light.
After the collision, centered on $v \,{=}\, 0$, 
energy is deposited throughout the region
between the two receding energy density maxima.
The energy density after the collision does not resemble the superposition
of two unmodified shocks, separating at the speed of light, plus small corrections.
In particular, the two receding maxima are moving outwards at 
less than the speed of light.
To elaborate on this point, Figure~\ref{fig:flux}
shows a contour plot of the energy flux $\mathcal S$ for positive $v$ and $z$.
The dashed curve shows the location of the maximum of the energy flux. 
The inverse slope of this curve, equal to the outward speed of the maximum,
is $V = 0.86$ at late times. The solid line shows the point
beyond which $\mathcal S/\mu^4 <  10^{-4}$, and has slope~1.  Evidently,
the leading disturbance from the collision moves outwards at the speed of light,
but the maxima in $\mathcal E$ and $\mathcal S$ move
significantly slower.

Figure~\ref{fig:pressures} plots the transverse and
longitudinal pressures at $z = 0$ and $z = 3/\mu$, as a function
of time.  At $z = 0$, the pressures increase dramatically during the collision, 
resulting in a system which is very anisotropic
and far from equilibrium.  
At $v = -0.23/\mu$, where $\mathcal P_{\|}$ has its maximum, 
it is roughly 5 times larger than
$\mathcal P_{\perp}$.  
At late times, the pressures asymptotically approach each other.
At $z = 3/\mu$,
the outgoing maximum in the energy density is located near $v = 4/\mu$.  There, 
$\mathcal P_{\|}$ is more than 3 times larger than $\mathcal P_{\perp}$.

The fluid/gravity correspondence \cite{Bhattacharyya:2008jc}
implies that at sufficiently late times the evolution of $T^{\mu \nu}$ will be described by hydrodynamics.
To test the validly of hydrodynamics,
Fig.~\ref{fig:pressures} also plots (as dashed lines) the pressures
$\mathcal P_{\perp}^{\rm hydro}$ and $\mathcal P_{\|}^{\rm hydro}$ 
predicted by the first-order viscous hydrodynamic constitutive relations
\cite{Baier:2007ix}.
At $z = 0$ the hydrodynamic constitutive relations hold within $15\%$
at time $v_{\rm hydro} = 2.4/\mu$, with improving accuracy thereafter.  
At $z = 3/\mu$, they hold within $15\%$ or better accuracy after
$v_{\rm hydro} = 2.1/\mu$.

At $z = 0$, where the flux $\mathcal S = 0$, the constitutive relations
imply that the difference between $\mathcal P_{\perp}^{\rm hydro}$ and $\mathcal P_{\|}^{\rm hydro}$ 
is purely due to viscous effects.   Fig.~\ref{fig:pressures} shows
that there is a large difference between
$\mathcal P_{\perp}$ and $\mathcal P_{\|}$
when hydrodynamics first becomes applicable,
implying that viscous effects are substantial.  

We have also examined the influence of second-order corrections in the hydrodynamic constitutive relations.
At $z = 0$, the second-order corrections only change $v_{\rm hydro}$
by about 1\%,
whereas at $z = 3/\mu$ their addition {\em increases} $v_{\rm hydro}$  by 20\%. Evidently, in front
of the receding maxima in $\mathcal E$ and $\mathcal S$, second-order corrections are large, 
the system is still far from equilibrium,
and agreement with the first-order constitutive relation
is largely fortuitous. 
 
Hydrodynamic simulations of heavy ion collisions at RHIC suggest that the
produced QGP
thermalizes in a time perhaps shorter than 1\,fm/c \cite{Heinz:2004pj}.
Crudely modeling the boosted gold nuclei
by our translationally invariant Gaussian shocks, 
for RHIC energies we estimate $\mu \sim 2.3$ GeV.
Our results from Fig.~\ref{fig:pressures}
then imply that the total time required for
apparent thermalization,
from when the Gaussian shocks start to overlap significantly to the onset
of validity of hydrodynamics,
is $\Delta v_{\rm tot} \sim 4/\mu \sim 0.35$ fm/c. 
Similar results for far-from-equilibrium relaxation times in SYM were also reported in
Ref.~\cite{Chesler:2009cy}.

We conclude by discussing the effect of the background energy density on our results.
In the absence of a collision, the presence of the background energy density
will cause a propagating shock to slowly attenuate in amplitude and eventually thermalize.
We have computed single shock propagation with the background energy
density used above,
and found that the shock attenuates in amplitude by 2.5\% in a time
$\Delta v = 1/T_{\rm bkgd}$. 
We have also studied the effect of increasing or decreasing the background energy density
by a factor of 1.5.  This results in a spacetime dependent $O(\delta)$ change in the stress
tensor and perturbs the thermalization time $v_{\rm hydro}$ (at $z = 0$) by 1\%.
Lastly, we have also studied an expanding 
sheet of plasma which is initially localized at $z =
0$ and surrounded by vacuum.  
The initial conditions consisted of $B = 0$, a Gaussian
profile in the energy density, and vanishing energy flux.
At late times the stress has two outward moving maxima,
just as it does for the colliding shocks.  Furthermore, the tails in the stress
move outwards at the speed of light whereas the maxima move
at a speed around $15\%$ less.
Consequently, we are confident that the deviation of the speed of the maxima
from 1 in
Figs.~\ref{fig:eden} and \ref{fig:flux} is not an artifact
caused by the background energy density.

\begin{acknowledgments}
The work of LY is supported by the U.S. Department
of Energy under Grant No.~DE-FG02-\-96ER\-40956.
The work of PC is supported by a Pappalardo Fellowship in Physics at MIT.
We are grateful to Scott Hughes, Paul Romatschke, and Ruben Rosales for useful discussions.
\end{acknowledgments}

\bibliographystyle{utphys}
\bibliography{refs}%
\end{document}